# A Computation-Efficient CNN System for High-Quality Brain Tumor Segmentation

**Yanming Sun and Chunyan Wang**

Department of Electrical and Computer Engineering, Concordia University, Montreal,

QC H3G 1M8, Canada

Corresponding author: Chunyan Wang (chunyan@ece.concordia.ca).

**ABSTRACT** Brain tumor diagnosis is an important issue in health care. Automated brain tumor segmentation can help timely diagnosis. The work presented in this paper is to propose a reliable high-quality system of Convolutional Neural Network (CNN) for brain tumor segmentation with a low computation requirement. The system consists of a CNN for the main processing for the segmentation, a pre-CNN block for data reduction and post-CNN refinement block. The unique CNN consists of 7 convolution layers involving only 108 kernels and 20308 trainable parameters. It is custom-designed, following the proposed paradigm of ASCNN (application specific CNN), to perform mono-modality and cross-modality feature extraction, tumor localization and pixel classification. Each layer fits the task assigned to it, by means of (i) appropriate normalization applied to its input data, (ii) correct convolution modes for the assigned task, and (iii) suitable nonlinear transformation to optimize the convolution results. In this specific design context, the number of kernels in each of the 7 layers is made to be just-sufficient for its task, instead of exponentially growing over the layers, to increase information density and to reduce randomness in the processing. The proposed activation function Full-ReLU helps to halve the number of kernels in convolution layers of high-pass filtering without degrading processing quality. A large number of experiments with BRATS2018 dataset have been conducted to measure the processing quality and reproducibility of the proposed system. The results demonstrate that the system reproduces reliably almost the same output to the same input after retraining. The mean dice scores for enhancing tumor, whole tumor and tumor core are 77.2%, 89.2% and 76.3%, respectively. Having 20308 trainable parameters, the system needs only 29.07G Flops to test a patient case, i.e., a very small fraction of what is required by the smallest CNN so far reported for the same segmentation. The simple structure and reliable high processing quality of the proposed system will facilitate its implementation and medical applications.

# 1. INTRODUCTION

Brain tumors pose a serious problem to human health, and brain tumor segmentation is a critical step for the diagnosis and treatment of the disease. Manual segmentation is very time-consuming and often causes delays. It is thus important to develop fully automated systems for brain tumor segmentation to facilitate timely diagnosis.

The automated brain tumor segmentation is a very challenging task. First of all, a human brain is very complex 3D structure. Brain tumor segmentation is to generate a 3D multi-class signal from four 3D MRI brain images, namely FLAIR, T1, T1c and T2. The four 3D images can be sliced into 4 sequences of 2D slices, and four 2D slices from the 4 sequences are shown in *Figure 1.1* (a)~(d), and the ground truth of the brain tumor segmentation corresponding to these slices is shown in *Figure 1.1*(e). From computer vision point of view, it is not easy to distinguish tumor tissues from healthy tissues, as the shapes, sizes, textures and locations of brain tumors are very different from patient to patient. Moreover, a high quality of brain tumor segmentation is required for meaningful applications in diagnosis. This quality requirement is not only in the voxel-wise precision to define whole tumor regions, but also in a fine classification of the voxels into three types of intra-tumoral structures, namely edema, non-enhancing (solid) core/necrotic (or fluid-filled) core and enhancing core [1].

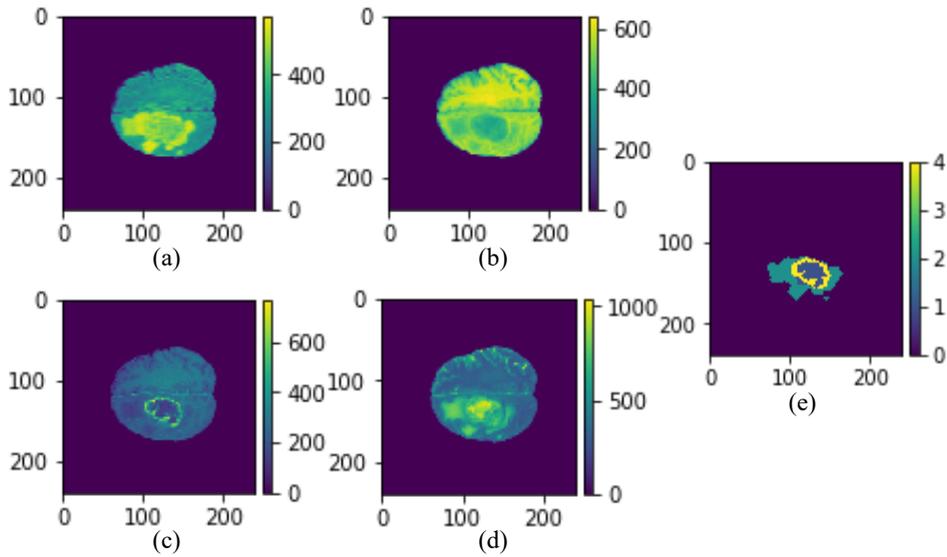

Figure 1.1 (a) FLAIR slice. (b) T1 slice. (c) T1c slice. (d) T2 slice. (e) Ground truth.

Brain tumor segmentation can be done by means of filtering and/or morphological operations. One of the segmentation methods is based on feature extracted by using Sobel high-pass filtering [2]. Tumor features can also be extracted by applying Gabor filters, and the filtered data are used to classify the pixels with extremely randomized trees [3]. Appearance- and context-based features of brain tumors can also be classified by using Extremely Randomized Trees [4]. The segmentation can also be performed by means of Random Forests model applied to the local texture and abnormality maps that are calculated from two MRI-sequences, namely FLAIR and T1c [5]. As healthy brains have certain degree of symmetry [6], brain tumor detection can also be done by analyzing the symmetry of brain images [7]. Implementations of these methods do not require large amount of computation. However, owing to the complexity of the brain tumor variations, it is difficult to precisely classify the pixels of intra-tumoral structures by simple filtering methods.

Convolutional Neural Network (CNN) is an effective approach to computer vision. The network structures, such as VGGNet [8], are widely used for various kinds of object recognition/detections. The filtering coefficients in a CNN are determined by means of progressive adjustment in a training process, without need to choose them manually. One can thus put a large number of kernels in each layer to handle intricate image features. If the training is done correctly, the filtering operations in a CNN can be well tunned to detect various features corresponding to the objects. Therefore, it has a good potential to deal with the variations of tumor features in brain images with a view to obtaining a good segmentation quality.

There are a good number of CNNs reported for brain tumor segmentation. Some use 2D convolutions, in which every convolution can be performed with multiple data channels, but each channel is a 2D map, and the others use 3D convolutions, requiring more computation to handle each data channel of 3D structure. Certain of them are related to the work presented in this paper and thus described in the next section.

In general, CNN systems require much more computation resources than those of deterministic filtering systems. With the development of IC technology, the limit of computation capacity has been extended. It should, however, be noted that the demands for integrating more functions with better functionality in mobile systems are growing endlessly at an even faster pace than the technology development. Hence, the computation resources can never be limitless, and continuously improving the computation efficiency, i.e., performing a given function with less computation, is a critical issue in designing CNN systems. Also, taking the cybersecurity problems into consideration, it is better to develop processing algorithms that are suitable for local computation, instead of centralized one. Moreover, for medical applications, reliable performance, fast processing, easy integration in computation resource restricted environment, are among the first items in the wishlist.

It should also be noted that the processing quality of a CNN depends not only on its computation structure, but also on its training. To train it appropriately, the number of training samples and the comprehensiveness of the target features in these samples should be sufficient, with respect to the number of parameters to be trained. If the quantity and/or quality is insufficient, which is often the case in reality, despite the measures such as data argumentation, the network can be trained with a lot of randomness, affecting the performance of the trained network. One may consider to reduce the number of kernels in the CNN to minimize the training randomness, while reducing the computation waste. The question will then be how to do it without degrading the processing quality.

The objective of the work presented in this paper is to design a reliable high-quality CNN system for brain tumor segmentation with low-computation requirement. The low computation is, on one hand, to facilitate implementation and applications of the system, and on the other hand, to obtain a high segmentation quality and good reproducibility. This high-quality is related to a high information density in the output channels of each convolution layer. For a given amount of information, the more channels, the lower the density. For a convolution layer, the higher information density in its input channels, the easier it can operate with a small number of kernels to produce its output channels of high information density, which facilitates more the processing in the succeeding layer. The design of the CNN system should get into such a positive circle.

This paper is to present how the objective has been achieved. For the processing quality not to be constrained by the low-computation requirement, a new design paradigm and new design components have been proposed, and applied in the design process, and remarkable results have been obtained.

The highlights of this work, constituting its contribution to the area of CNN development, are presented as follows.

- New design paradigm of Application-Specific CNN (ASCNN). An ASCNN is not a modified or extended version of a commonly used structure, e.g., U-net [9], or combination of such structures, but a network specifically built with basic convolution modules for a given task, e.g., a detection of specific targets. This paradigm makes it possible to design a reliable CNN system of very simple structure to do the task with a very small fraction of the computation required by others, without sacrificing the processing quality.

- New activation function Full-ReLU. Though it is designated to perform a nonlinear transformation, Full-ReLU can also effectively minimize the number of the kernels required in a convolution layer for high-pass filtering without degrading processing quality.

- New CNN architecture custom-designed for brain image segmentation. It has 7 convolution layers, utilizing, in total, 108 3×3 convolution kernels, i.e., less than 16 kernels per layer in average. Each of the processing stages in the CNN system is designed with computation elements that specifically suit its input data and help to produce the output data of high information density. The segmentation quality of the system is better than most of the CNNs reported for brain tumor segmentation.

- Design for reproducibility and assessment of performance consistency. A reliable system must be able to deliver its results in a consistent manner and its results should be reproduceable, which is critically important for its medical applications. The proposed CNN has been designed with a view to minimizing the randomness in its computation structure and in the training process, which leads to a very good performance consistency.

The paper consists of 5 sections. The related work is presented in Section II. The detailed description of the system design is found in Section III, and that of the training and testing is presented in Section IV. A conclusion is presented in Section V.

## 2. RELATED WORK

The related work is about CNNs for brain tumor segmentation. In this subject domain, U-net based CNNs are, no doubt, the most commonly seen in literature, but there are also non-U-net CNNs. From convolution operation point view, some are 2D CNNs and the others are 3D. They are presented in this subsection.

2D U-net, having some similarity to a network called FCN [10], has been widely used in medical image segmentations [9]. The main frame of the U-net consists of a VGG-like contracting path, in which the number of convolution kernels increases exponential, and an expanding path, in which copies of the feature data generated in the contracting path are included by concatenation in the convolutions. This main frame is found, with some variations, in many CNNs for brain tumor segmentation. For example, standard convolution blocks in the original U-net are replaced by inception blocks [11] to enhance the feature extraction. U-net structure can also be used as a basic unit, and multiple units are stacked to seek a better processing quality [12]. In DRINet [13], standard convolution blocks in the first layers are replaced by dense connection blocks, and those in the last layers are replaced by residual inception blocks, whereas the copy/crop part (also called long skip connection) is not used. The network presented in [14] was designed to have an adaptive feature recombination and recalibration to improve the segmentation quality.

Other forms of 2D CNN have also been adopted for brain tumor segmentation. A deep residual dilate network with middle supervision is used to segment multi-modal brain tumor images [15]. The method based on multi-cascaded convolutional neural network (MCCNN) and fully connected conditional random fields (CRFs) is also reported and the patches obtained from axial, coronal, and sagittal views to respectively train three segmentation models, and then combine them to obtain the final segmentation result [16].

Since brain images are represented by 3D data, 3D CNNs are used for brain tumor segmentation. It should be noted that, in a 3D CNN, each input data channel is a 3D body, instead of 2D slice. The 3D dense connectivity structure with atrous convolutional feature pyramid [17] is a 3D CNN for brain tumor segmentation, and it employs a deep supervision mechanism to promote training. A dual-force training scheme is also reported to train two 3D CNNs, named DF-U-Net and DF-MLDeepMedic, to promote the quality of multi-level features learnt from deep models [18]. Besides training methods, one can incorporate multiple 3D CNNs for a better performance. DF-MLDeepMedic involves 2 pathways of DeepMedic to make use of multi-level information for more accurate segmentation [18]. There is also a CNN structure of several 3D U-nets grouped into 3 stages for generating constraints, fusion under constraints, and final segmentation, respectively [19].

It should be mentioned that, a VGG-like contracting path with its exponentially growing convolution kernel number is a staple element in all the CNNs reported for brain tumor segmentation, no matter they are U-net based or not. It requires, however, a large number of trainable parameters to fill up the kernels. The number can be counted from millions to billions. It may still be considered normal, and the training/testing can be performed quickly with powerful GPUs, if the CNN operates in a stand-alone manner. It is, however, excessive if such a network needs to be implemented in a mobile device in order to operate within a sophisticate graphics application. The design paradigm of ASCNN is thus proposed aiming at this problem, and it is applied in the design of the CNN system presented in the following section.

## 3. PROPOSED SYSTEM

As mentioned previously, each patient case, as the input of a segmentation system, is represented by four 3D MRI images of 4 different modality, namely FLAIR, T1, T1c and T2. As each of the 3D images is sliced into 2D slices, a voxel in a 3D image becomes a pixel in a 2D slice. In the proposed CNN system, the brain tumor segmentation of 3D images is done by segmenting these 2D slices.

The proposed system is designed in such a way that the main data processing for the segmentation is performed in a CNN and there are also pre-CNN and post-CNN blocks to reduce the computation burden of the CNN, aiming at a better computation efficiency of the CNN. Moreover, by doing so, the system will have a higher degree of determinism, less randomness, and consequently a better reproducibility. The overview of the proposed system is illustrated in *Figure 3.1*. The functions of the three parts are as follows.

- Pre-CNN. It is to reduce the volume of input data applied to the succeeding CNN.
- CNN. It is designed specifically to perform brain tumor segmentation. It has a unique structure of convolutions and employs a new activation function in the convolution layers for feature extraction.
- Post-CNN. It is to detect and to remove possible false positive pixels.

This section is organized as follows. The pre-CNN and post-CNN are described in Subsections 3.1 and 3.3, respectively. Subsection 3.2 is dedicated to the presentation of the design of the CNN, the main part of the proposed system. The new activation function Full-ReLU is described in the subsection 3.2.1.

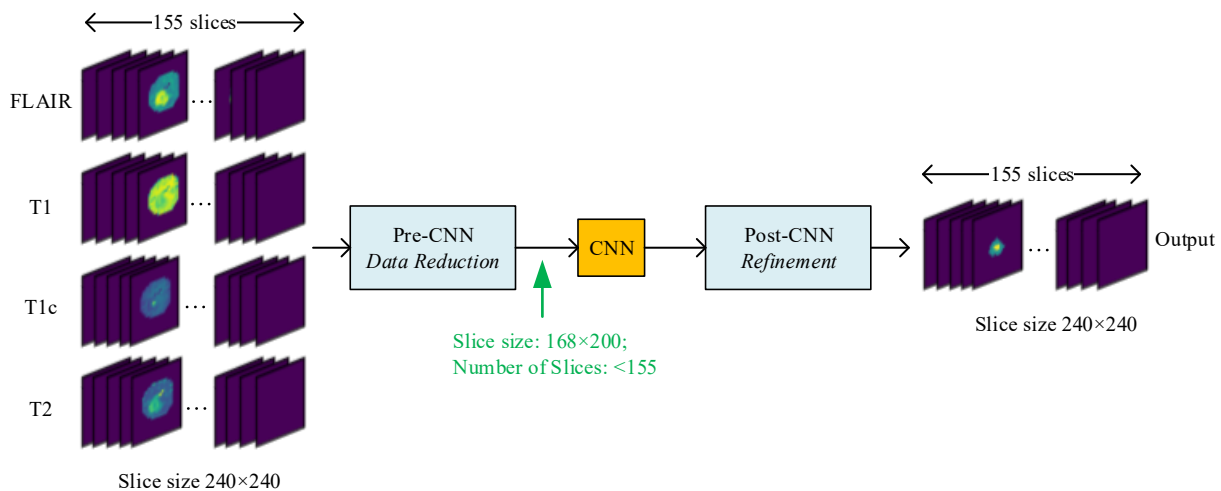

Figure 3.1 Overview of the proposed 3-block system. There are four 3D images namely FLAIR, T1, T1c and T2 in each patient case and the size of each 3D image is 240×240×155.

## 3.1 PRE-CNN – DATA REDUCTION

The dimensions of the input 3D images from commonly used datasets, such as BRATS2017 and BRATS2018, are 240×240×155, resulting from a post-acquisition registration. The pre-CNN is to reduce the data volume of the 3D images to facilitate the computations in the CNN. To be specific, it is done by detecting and removing the following 3 kinds of tumor-free slices in brain images.

(i) In the sequence of 155 slices in each 3D image, brain tumor areas are seldom found in the slices located at the 2 ends of the sequence. The slices of this kind can be identified by their very high percentage of background pixels. Two examples of such slices are shown in *Figure 3.2*.

(ii) The slices of the second kind are of healthy section of a brain. Healthy brains have natural left-right symmetry, which is reflected to the upper-lower symmetry in brain images [20, 21], as an example shown in *Figure 3.3* (a). A tumor causes an asymmetry, as a slice with tumor shown in *Figure 3.3* (b). Thus, a tumor-free slice of this kind can be identified by a high degree of upper-lower symmetry of the pixel data inside the brain area.

(iii) Slices of incomplete appearance of brain areas, as examples shown in *Figure 3.4* (a) and (c), are produced due to imperfection in medical image acquisition. They are also considered as tumor-free slices, as it is very rare to find brain tumors in the locations of this kind of slices. Because of the incomplete appearance of brain areas, such a slice has a significant asymmetry between the outlines of the upper and lower halves, as shown in *Figure 3.4* (b) and (d). Hence one can identify it by calculating the degree of asymmetry of the outlines, ignoring the contents inside the areas.

There are various methods to measure the degree of symmetry/asymmetry of the upper and lower halves of brain area contents or brain area outlines. Structural similarity (SSIM) [22] is used in this design to identify tumor-free slices of the second and third kinds. If $x$ and $y$ are 2 sets of data, their SSIM is calculated as follows:

$$SSIM(\boldsymbol{x},\boldsymbol{y}) = [l(\boldsymbol{x},\boldsymbol{y})]^\alpha \cdot [c(\boldsymbol{x},\boldsymbol{y})]^\beta \cdot [s(\boldsymbol{x},\boldsymbol{y})]^\gamma \tag{1}$$

where

$$l(\boldsymbol{x},\boldsymbol{y}) = \frac{2\mu_x\mu_y + C_1}{\mu_x^2 + \mu_y^2 + C_1}, \quad c(\boldsymbol{x},\boldsymbol{y}) = \frac{2\sigma_x\sigma_y + C_2}{\sigma_x^2 + \sigma_y^2 + C_2}, \quad s(\boldsymbol{x},\boldsymbol{y}) = \frac{\sigma_{xy} + C_3}{\sigma_x\sigma_y + C_3}$$

$(\mu_x, \mu_y)$ denote mean values of $x$ and $y$, $(\sigma_x, \sigma_y)$ are the standard deviations, $\sigma_{xy}$ is the correlation coefficient, $(\alpha, \beta, \gamma)$ denote the relative importance of the three parts, and $(C_1, C_2, C_3)$ are small non-zero constants. The default values for $(\alpha, \beta, \gamma)$ are (1, 1, 1).

Applying the method described above to the samples from BRATS2018, one can find that the number of tumor-free slices per case is between 13 and 43, i.e., 8.4% ~28% of the 155 slices, which is not negligible. Combining it with the removal of the excessive margins in each slice, a reduction of more than 50% data volume can be achieved, which reduces significantly the computation load in the succeeding CNN stage.

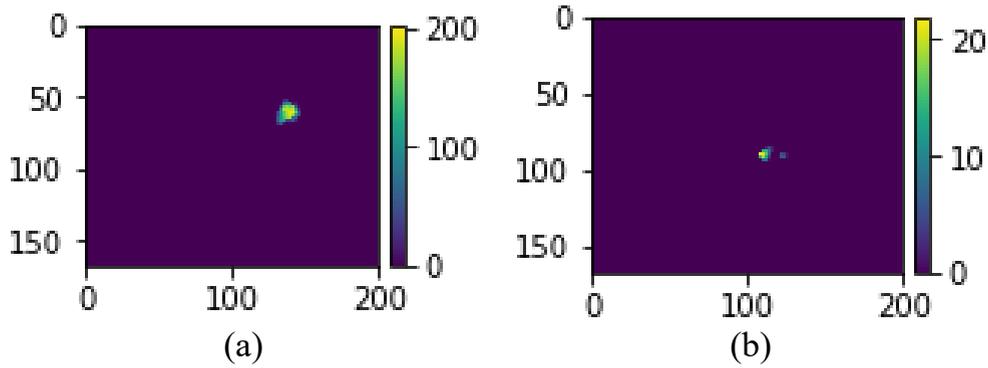

Figure 3.2 Examples of slices with high percentage of background pixels.

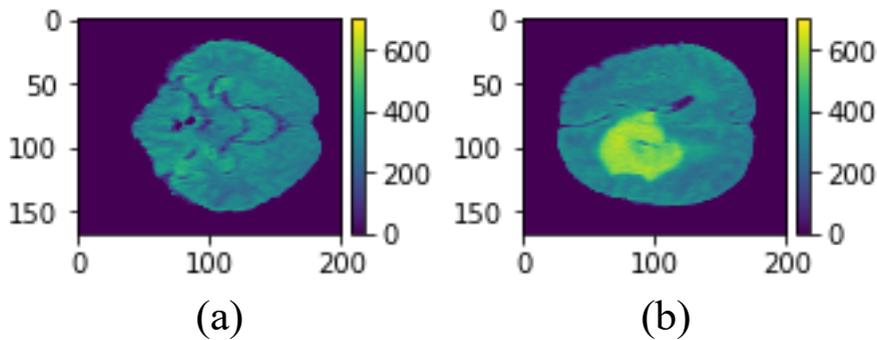

Figure 3.3 (a) An example of a FLAIR slice without tumor. The SSIM value of the upper and lower halves is 0.308. (b) An example of a FLAIR slice with tumor. The patterns in the upper and lower halves do not mirror themselves. Its SSIM value is 0.174.

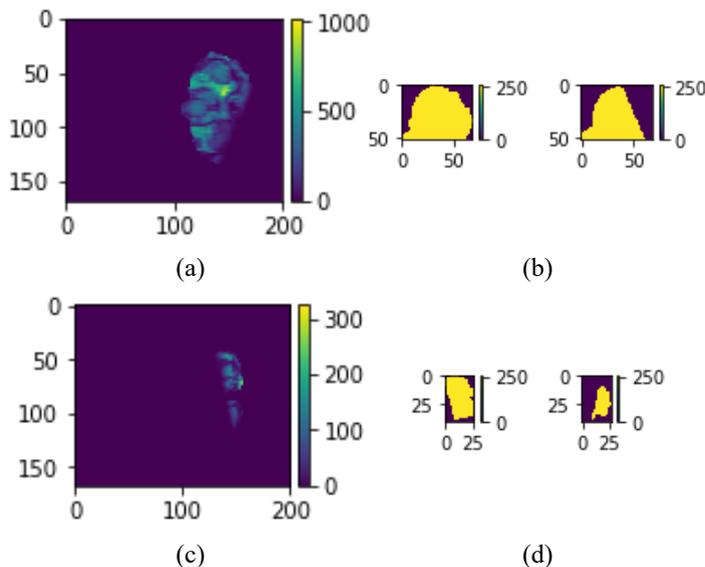

Figure 3.4 (a) (c) Examples of slices with incomplete appearance of brain areas. (b) (d) Upper and lower halves of the binary image of (a) and (c) that highlights the outlines of the brain areas.

## *3.2 CNN*

As the proposed CNN is to perform the main segmentation function, its processing quality determines the overall performance of the entire system. In order not to compromise the processing quality while minimizing the amount of computation, one needs to custom-design the CNN, instead of simply adopting and modifying an existing network. By doing so, all the computation elements are made necessary and just-sufficient to process the input data for the specific brain segmentation task. Also, a new activation function, referred to as Full-ReLU, is used in some layers, which helps to make the computation more efficient.

The new activation function Full-Relu is proposed in the subsection 3.2.1. The design of the unique CNN, having a number of particular design components, is presented in the subsection 3.2.2.

### 3.2.1 FULL-RELU – A NEW ACTIVATION FUNCTION

An activation function is used to implement a nonlinear transformation in a convolution layer. ReLU may be the most commonly used activation function in CNN designs. It is, however, not flawless. In case of high-pass filtering, a single convolution operation is able to generate a set of positive and negative elements, representing the signal variations in the two opposite directions. If ReLU is applied to these elements, the negative ones will be eliminated, resulting in information loss. In order to compensate for such a loss, one can double the number of kernels in the convolution layer, attempting to catch the information of both directions, at the expense of more computation and, in some cases, risk of adding noisy components.

The new activation function is proposed in order not to lose the information carried by the negative elements in the situation described above so that the number of kernels can be halved in such a convolution layer.

It is to produce, from one set of data $[X]$, 2 sets, namely $[X_p]$ and $[X_n]$. Its characteristics are illustrated in *Figure 3.5* and mathematically expressed as follows.

$$\begin{cases} x_{p\,ij} = max(0, x_{ij}) \\ x_{n\,ij} = |min(0, x_{ij})| \end{cases} \tag{2}$$

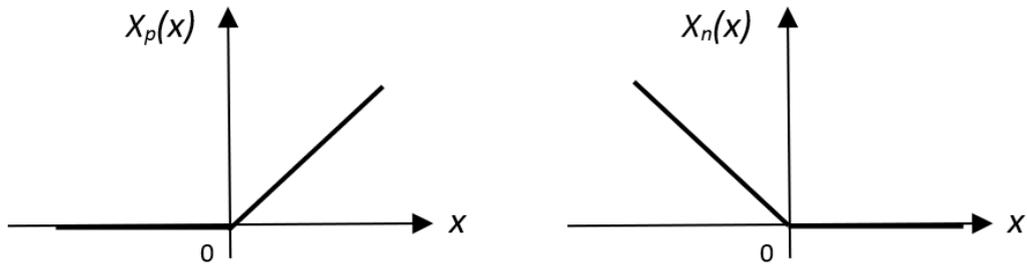

Figure 3.5 Characteristics of Full-ReLU.

If [X] is a feature map produced by a convolution, its elements will be sorted and preserved in the 2 different feature maps [$X_p$] and [$X_n$], without mix-up. We can say that the proposed activation function not only provides a nonlinear transformation, but also groups the feature data in a meaningfull manner.

The new activation function is referred to as Full-ReLU, as it checks the signal magnitudes, like the function of full-wave rectifier. It should, however, be underlined that Full-ReLU sorts the data by their signs, unlike full-wave rectifier.

It should be noted that Full-ReLU is not to replace the conventional ReLU everywhere in a CNN. It is suitable in stages of feature extraction by high-pass filtering, or other places where elements having negative values need to be preserved separately.

### 3.2.2 CNN DESIGN

The proposed CNN is specifically designed to receive 2D brain image slices of the 4 modalities and to classify the pixels into 4 classes, i.e., whole tumor (WT), tumor core (TC), enhancing tumor (ET), and background. The convolution layers are to perform (a) feature extraction from the input of 4 modalities, (b) tumor localization, and (c) tumor pixel classification. With this plan, each layer is assigned a particular task. The computation capacity of each layer is made to just sufficient for the assigned task, no more no less. This is, however, in the context of the following 3 issues.

- The input data are well prepared, e.g., normalized appropriately considering the statistical natures of the data.
- The convolution performed in the layer is appropriate for its input data and assigned task.
- The convolved data are well handled to maximize their utilization, minimizing information loss.

The detailed block diagram of the proposed CNN is shown in *Figure 3.6*. The model hyperparameters in its configuration, presented in *Table 3.1*, are very different from the other CNNs. The most important character of this CNN is its remarkable simplicity: 108 convolution kernels in total, used in the 7 convolution layers. The number of kernels per convolution layer does not increase exponentially from layer to layer. For this extremely simple CNN to produce the results of a good quality, each convolution layer is designed, in view of the 3 above-mentioned issues. To this end, the following points should be well taken care of in the design.

- *Modes of convolution*. There are 3 modes, namely standard convolution, depthwise convolution and group convolution, differing themselves on how each convolution kernel is applied to the input channels. A kernel of standard convolution is applied to all the input channels at once, whereas that of depthwise convolution is applied to only one at a time, and that of group convolution is to several at a time.
- *Normalization*. It is to uniform the data distributions of the different channels so that they are prepared for the succeeding convolution. The data preparation can make significant difference in the processing quality.
- *Number of kernels in each convolution*. It determines the capability to extract varieties of features and to accommodate the feature data. It should be carefully tailored to be just sufficiently large to handle the complex features in the input. An excessively large number of kernels will result in not only computation waste and longer training/testing time, but also lower information density in data channels, more randomness and noise components in the system.
- *Activation function*. In general, it helps to implement a nonlinear transformation in a convolution layer. The special activation function Full-ReLU can also help to better handle the convolved data in order to reduce data loss, if it is used appropriately.

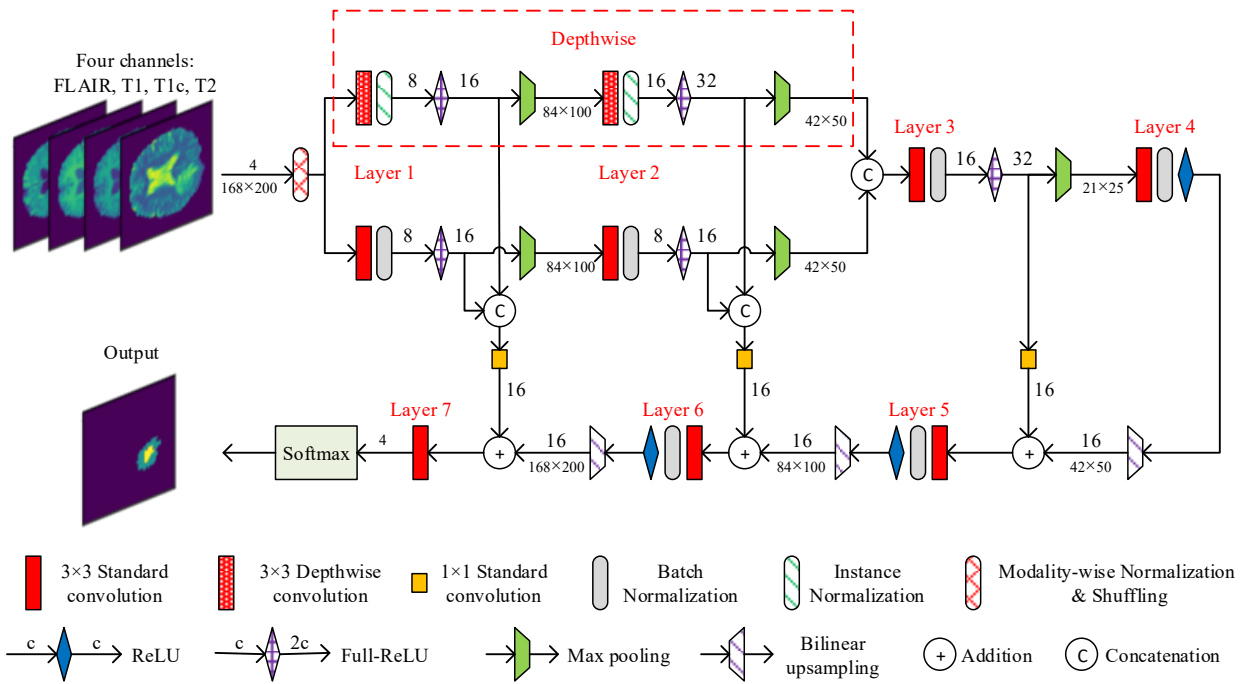

Figure 3.6 Detailed block diagram of the proposed CNN.

Table 3.1 Details of the proposed CNN Configuration

| Layer# | Input channels# | Kernels# | Activation function | Output channels# | Size of input maps | Size of output maps | Parameters# |
|---|---|---|---|---|---|---|---|
| 1 | 4 | 8, 8 (D, S) * | Full-ReLU | 16, 16 | 168×200 | 84×100 | 376 |
| 2 | 16, 16 | 16, 8 (D, S) * | Full-ReLU | 32, 16 | 84×100 | 42×50 | 1320 |
| 3 | 48 | 16 | Full-ReLU | 32 | 42×50 | 21×25 | 6928 |
| 4 | 32 | 16 | ReLU | 16 | 21×25 | 42×50 | 4624 |
| 5 | 16 | 16 | ReLU | 16 | 42×50 | 84×100 | 2320 |
| 6 | 16 | 16 | ReLU | 16 | 84×100 | 168×200 | 2320 |
| 7 | 16 | 4 | ReLU | 4 | 168×200 | 168×200 | 580 |
| | | | Three 1×1 Convolutions, 528+784+528 | | | | 1840 |
| | | | Total | | | | 20308 |

*D: Depthwise convolution kernels for mono-modality feature extraction in the first and second layers
*S: Standard convolution kernels for cross-modality feature extraction in the first and second layers

*3.2.2.1 Layers 1 and 2 – Feature extraction by 2-path convolutions*

First of all, the input data need to be prepared, by normalization, for the very first convolution. As the brain images may not be acquired under the exactly same condition, the intensity range of the brain image data may vary from patient to patient, or from one modality to another [23, 24]. This normalization is specifically "modality-wise", performed to each of the four 3D images of a patient case. For example, if there are *N* slices in a 3D image of FLAIR modality, the data of all the *N* slices are normalized with their mean and standard deviation. This normalization uniforms the data range and gray level distribution of all the input channels of the network, facilitating the filtering operations in the convolution layers.

The first 2 layers are to extract basic features elements from the normalized data of the input brain image slices. The quality of the features generated in this part can impact the overall performance of the network. To extract features comprehensively and also specifically, the input data are processed in parallel in the 2 paths, as shown in *Figure 3.6*. In each of the paths, the convolution mode, the normalization, the activation function, and the number of kernels are determined, taking very carefully the input data into consideration.

The data of the 4 different modalities are acquired with different emphases on different lesion areas of a specific brain. Each of them contains enhanced features of particular intra-tumoral structures. To make good use of the data of each of the 4 modalities, 2 successive depthwise convolutions, represented by red-white-polka-dot rectangles in *Figure 3.6*, are applied individually to each of the 4 slices, to generate specifically 2D feature maps of mono-modality. Also, the successive depthwise convolutions permit to produce, from the same input slice, feature data of different filtering orders.

Two kernels are applied, in a depthwise convolution of the first layer, to each of the 4 input slices, to detect features of 2 different orientations, and with the help of Full-ReLU, this convolution produces 4 feature maps from that input slice. In total, 16 maps of the first-order mono-modality features are produced from the 4 input slices. Each of the 16 is then convolved with a single kernel in the second layer, and again by means of Full-ReLU, there will be 32 second-order mono-modality feature maps.

The data of each feature map generated by a depthwise convolution are normalized with the mean and standard deviation of that particular map. It is referred to as channel-wise normalization, as in 2D convolution, each input/output channel is a 2D map. It is also called "instance" normalization in some literature.

It should be mentioned that, though the 4 slices of different modality have different emphasis on brain image features, they represent the same brain section of the same patient and there is a correlation among them. Hence, standard convolutions are also used in the first 2 layers of the proposed CNN to generate, in a comprehensive manner, 2D feature maps of cross-modality. They are performed in parallel with the depthwise convolutions as shown in *Figure 3.6*. Eight kernels used in the first convolution and 8 in the second one. Full-ReLU is used after each convolution, and the 8 kernels result in 16 feature maps.

In summary, the total number of kernels used in the first layer is 16 and that in the second layer is 24, generating 32 and 48 feature maps, respectively. These feature maps are used, in the following layers, not only to localize brain tumors but also to classify the pixels.

It should be mentioned that a pooling operation is performed to each of the 2D feature maps, as shown in *Figure 3.6*. It is to "zoom out" the 2D map, so that the effective size of neighborhood in the succeeding convolution will be larger than the size of the kernels of 3×3 pixels. By doing so, the extracted feature information can get more comprehensive and more condensed over the layers.

*3.2.2.2 Layers 3 and 4 – Tumor localization, 16 kernels per layer*

The overall function of the $3^{rd}$ and $4^{th}$ layers is to generate, from the cross-modality and mono-modality feature data, a series of maps, in each of which the pixels representing tumor features are distinguished from the rest of the population. These maps are used, all together, to indicate approximate locations of tumor candidates, ignoring some tumor details due to the low resolution. It is done in 2 steps: the transformation of feature data by the $3^{rd}$ layer and the generation of low-resolution location maps by the $4^{th}$ layer.

Standard convolutions are performed in the 2 layers so that the output is generated comprehensively from all the input channels. A regular batch normalization is applied after each convolution.

Like the $1^{st}$ and $2^{nd}$ layers, there is a pooling operation after the convolution in the $3^{rd}$ layer. Thus, the data map is "zoom-outed" one more time. Hence, in the $4^{th}$ layer, the value of each pixel is calculated based on the feature data originated from a very large neighborhood, which is necessary as tumor areas can be identified by certain textures/patterns from a macroscopic point of view, not image details in small windows.

It should, however, be noted that the 3 pooling operations reduce considerably the signal resolution. Therefore, the data maps applied to the 4$^{th}$ layer carry low-resolution image features, representing very "coarse" brain tissue textures.

The number of kernels in each of the 2 layers is determined, as mentioned previously, according to the complexity of the input data and the assigned task. Brain tumors can have various appearances in the background of brain images. The variations in the background are, however, much more limited, with respect to many other cases, e.g., image samples from ImageNet. Thus, the amount of computation allocated in the 3$^{rd}$ and 4$^{th}$ layers can be much less than that in many other object detections, and a modest number of kernels will be sufficient to handle the data variations. Moreover, a smaller number of kernels results in a smaller number of output data maps, increasing the information density in each map.

The convolution in the 3$^{rd}$ layer uses only 16 kernels, and by means of Full-ReLU, 32 feature maps are generated to be applied to the 4$^{th}$ layer. There are also 16 kernels in the 4th layer to generate 16 maps that present, collectively, the tumor location information. A regular ReLU, instead of Full-ReLU, is used in the 4$^{th}$ layer, as the elements in the16 maps are high-order filtering results and unlikely to have negative values.

In summary, in the 3$^{rd}$ and 4$^{th}$ layers, the high-density feature maps are received and processed in order to generate only 16 data maps. They are expected to carry high-quality high-density information of tumor locations, facilitating the classification in the following 3 layers.

*3.2.2.3 Layers 5, 6, and 7 – Fine classification of pixels*

The overall function in the 5$^{th}$, 6$^{th}$ and 7$^{th}$ layers combined, illustrated in the lower section of the structure shown in *Figure 3.6*, is to generate a 4-class map of the original size, from all the data maps generated successively in the first 4 layers. This part of the network share 2 common points, in principle, with the expanding path of U-net: i) As the map size has been reduced by the pooling operations, upsampling operations are used to recover gradually the map size. ii) As the fine classification of pixels needs detailed features in tumor areas, the feature maps produced in the 3$^{rd}$, 2$^{nd}$, and 1$^{st}$ layers are introduced, one after another, to the convolutions of the last 3 layers. However, the 5$^{th}$, 6$^{th}$ and 7$^{th}$ layers have the following characters differing themselves from many U-nets found in literature.

- Bilinear upsampling is used to expand the map size. Compared with upsampling with trainable parameters, such as deconvolution or transposed convolution, it requires much less computation, without dependency on training, without introducing any randomness, although it does not create image details.
- Data modulation by means of a weighted addition. A 1×1 convolution is used to transform local feature maps, generated in one of the first 3 layers, into modulating feature maps that are then added, not concatenated, to the upsampled data, as shown in *Figure 3.6*. In this way, the upsampled data are modulated with the local feature data. It results in an enhancement of the feature data in tumor areas, which implies an attenuation of those in tumor-free areas.

- Neither normalization nor ReLU/Full-ReLU applied to the modulating feature maps. The 1×1 convolution is a linear operation combining its input maps with trainable weights. If the training is done appropriately, the distribution of its output, i.e., the modulating feature maps, will be appropriate for the succeeding modulation. Hence, neither normalization nor nonlinear manipulation is needed.
- Neither normalization nor ReLU/Full-ReLU applied to the output data of the last convolution. As mentioned previously, normalization is, in general, a data preparation for the succeeding convolution layer. No normalization is needed at this point. Also, the values of these output data are within the range of the succeeding Softmax, applying ReLU/Full-ReLU could result in data loss or unnecessary data separation.
- Sixteen kernels per layers. The $5^{th}$ layer has only 16 input channels, or 16 input maps. The information density in these maps is very high, with respect to a case of, e.g., 256 maps, which facilitates the processing in the $5^{th} \sim 7^{th}$ layers. With such high-quality input data, no large convolution layers are needed in the stage of fine classification, which has been confirmed in the system tests.

*3.2.2.4 Summary of the CNN design*

In the proposed system, the segmentation is mainly done by the CNN, and it is custom-design to achieve the best computation efficiency. To this end, each layer is assigned a particular task, and designed to fit it. In particular, the following elements in each layer are carefully determined according to its input data and the task assigned to it.

- Input data normalization. A modality-wise normalization is applied to the very input image slices, instance normalization applied to the data produced by a depthwise convolution, and no normalization is applied to certain kinds of data.
- Convolution mode. For example, depthwise convolution is used to extract mono-modality features and standard convolution for cross-modality features.
- Nonlinear transformation of convolved data. Full-ReLU, the new activation function proposed in this paper, is applied in the high-pass filtering layers to halve the number of kernels, and regular ReLU in the other layers, except the last one.

In such a setting, the number of kernels in each layer is minimized to be just-sufficient for a good processing quality. With respect to many networks reported for the same task, this CNN can be expected to result in a) better training of the filter coefficients and less dependency on training samples, b) higher information density in each output map, c) less room for randomness and computation inefficiency, and d) easier implementation of the CNN and integration in various systems.

### *3.3 POST-CNN – REFINEMENT*

After the classification by the CNN, the post-CNN block is placed to identify the pixels that are falsely classified as positive ones. The identification is based on the fact that a brain tumor and its enhanced tumor core, if it exists, are 3D objects, and the area of each of them must be found in a certain number of consecutive slices.

The thickness of a detectable whole tumor is considered to be at least 1/20 of the diameter of a brain. If a 3D brain image consists of 155 slices, this thickness corresponds to at least seven consecutive slices. If a whole tumor area appears in fewer than seven consecutive slices, the pixels in this area are likely falsely classified, and will be re-classified as tumor-free pixels.

The identification of the false-positive pixels of enhancing tumor core is based on a similar principle that tumor cores have their own minimum size limit. Also, it is common sense that the minimum size of a tumor core is slightly smaller than that of a whole tumor. If a predicted tumor core area appears in fewer than six consecutive slices, instead of seven, the pixels in the area will be considered false-positive and then be reclassified as non-enhancing tumor pixels.

By applying the principles mentioned above, the post-CNN block improves the precision of segmentation without adding a perceivable amount of computation.

In conclusion, the CNN, which is designed specifically based on the characters of the brain images, combined with the pre- and post- CNN blocks, performs the brain tumor segmentation precisely and efficiently with a very low computation cost.

## 4. PERFORMANCE EVALUATION

The proposed system for brain tumor segmentation has been trained and tested with a commonly used dataset, namely BRATS2018, for its performance evaluation. In Subsection 4.1, the datasets, tools and training details are presented. The performance evaluation process and results are found in Subsection 4.2, including the assessment of the reproducibility and consistency in performance in the subsection 4.2.2, the ablation study in the subsection 4.2.3, and the performance comparison in the subsection 4.2.4.

## 4.1 DATASET AND TRAINING PROCESS

### 4.1.1 DATASET AND TOOLS

The dataset BRATS2018 includes all of the samples from BRATS2017 dataset. There are 285 patient cases, including 75 LGG cases, in the training pool. It also provides 66 cases without ground truth for testing, and to evaluate the segmentation results, one can use CBICA Image Processing Portal [25], an online evaluation platform, where the assessment is a standard process with data from the Cancer Imaging Archive [26, 27].

As the number of patient cases from the dataset is very limited, data augmentation has been performed to have a decent training process. It was done by up-down and left-right flipping each slice in all the 3D brain images in the training set. Moreover, in order to include slices of different texture pattern in each batch, the slices in the training set have been sorted by shuffling them randomly.

Python and Tensorflow have been used in developing & testing the system, and a NVIDIA P100 Pascal GPU with 12GB HBM2 memory has been used to run the programs.

### 4.1.2 TRAINING DETAILS

In the proposed CNN, there are 20308 parameters to be determined by means of training process. A number of elements are critical for the quality of the training:

- Batch size and the number of training epochs. The batch size is chosen to be 100, and the training process is completed after 50 epochs.
- Learning rate. Cosine Decay [28] is chosen to make the learning rate variable from 0.01 to $1\times10^{-6}$. In this way, the system loss is reduced coarsely and quickly during the first epochs and is then adjusted finely in the last epochs.
- Loss function. The loss function is chosen to be Cross Entropy [29].
- Optimizer. The optimizer is chosen to be Adaptive Moment Estimation (Adam) [30].
- Initialization. The initial weights are chosen to be truncated normal distribution with 0.1 standard deviation, and the initial biases are chosen to be 0.1.

## 4.2 TESTING AND PERFORMANCE EVALUATION

After determining the values of the trainable parameters in the CNN by the training process, the proposed system has been tested by using BRATS2018 test (validation) set. All the data about the performance of the proposed system are the evaluation results by CBICA.

### 4.2.1 PERFORMANCE METRICS

There are four commonly used metrics to evaluate the segmentation quality, namely Dice score (*Dice*), Sensitivity (*Sens*), Specificity (*Spec*) and Hausdorff95 distance (*Haus*) [1]. The first three metrics are defined as follows.

$$Dice(P_1, T_1) = \frac{P_1 \wedge T_1}{(P_1 + T_1)/2} \qquad (3)$$

$$Sens(P_1, T_1) = \frac{P_1 \wedge T_1}{T_1} \qquad (4)$$

$$Spec(P_0, T_0) = \frac{P_0 \wedge T_0}{T_0} \qquad (5)$$

where $P_0$ and $P_1$ are the predicted results, indicating the number of voxels in the tumor-free regions and that in the tumor regions, respectively, whereas $T_0$ and $T_1$ are those in the ground truth. In addition, Haus is used to indicate the distance between the predicted tumor boundaries and those of the ground truth [1].

As the tumor voxels take a very small portion of space in a 3D brain image, we have

$$P_1 \ll P_0, \qquad T_1 \ll T_0, \qquad P_0 \approx T_0 \approx P_0 \wedge T_0, \qquad Spec(P_0, T_0) \approx 1$$

As $Spec(P_0, T_0)$ is almost equal to unit in most of brain tumor segmentation cases, it does not indicate sensitively a difference in identification of true/false negative voxels.

To better evaluate the quality of the segmentation in different aspects, the metrics of False Discovery Rate (*FDR*) [31] and False Negative Rate (*FNR*, or miss rate) [32] of the proposed system have also been measured. False Discovery Rate (*FDR*) is the ratio of the number of false positive voxels to the total number of predicted positive voxels, indicating how many voxels are falsely predicted to be positive, whereas *FNR*, or miss rate, is the ratio of the number of false negative voxels to the total number of positive voxels in the ground truth. They are expressed as follows.

$$FDR(P_1, T_1) = \frac{P_1 - (P_1 \wedge T_1)}{P_1} = 1 - \frac{1}{2/Dice(P_1, T_1) - 1/Sens(P_1, T_1)} \qquad (6)$$

$$FNR(P_1, T_1) = \frac{T_1 - (P_1 \wedge T_1)}{T_1} = 1 - Sens(P_1, T_1) \qquad (7)$$

The performance metrics of a CNN system also include measures of the computation volume required to achieve the processing quality, as it is related to the computation efficiency, the feasibility of system implementations, and the range of applications. The number of parameters in the CNN is an important indicator of the computation volume in both training and testing process. The number of floating-point operations (FLOPs) required to complete a test for one patient case is related to the applications of the system, as it determines where the system can be installed and how fast the process will be.

## 4.2.2 REPRODUCIBILITY AND CONSISTENCY IN PERFORMANCE

A functional system should be able to reproduce, from the same inputs, the same, or very similar, results every time after the system is reset/retrained. The degree of reproducibility reflects the degree of the reliability and confidence of the results. The reproducibility is ultra-important for CNN applications in medical image processing. Therefore, the assessment of the degree of reproducibility should be part of the validation of the results.

The reproducibility is an underline issue in a neural network, due to varieties of randomness in the computing process for training/testing. In general, the more complex the network, the more randomness. The proposed CNN system is designed to minimize the randomness and thus expected to have a good reproducibility.

Ten experiments have been conducted to assess the reproducibility. Each experiment has been done by (i) training the system from the initial state and (ii) testing all the 66 testing cases in the testing pool of BRATS2018 and generating 66 sets of scores, each of which includes the 3 Dice scores for ET, WT and TC.

From the 66 sets of the 3 Dice scores generated in each experiment, the statistical feature data, i.e., the mean, median and mode values of the scores, have been calculated. The statistical data from all the 10 experiments are illustrated in *Table 4.1*. Each column in the table contains 10 statistical feature data of the same kind, e.g., the mean values of ET Dice scores, generated in the 10 experiments, and the values are very close to each other. These data demonstrate that the functionality of the system is statistically consistent after any of the 10 training processes.

It is also observed, from the statistical data presented in *Table 4.1*, that the median values of all the 3 Dice scores are, consistently in all the 10 experiments, significantly higher than the mean values, indicating that most Dice scores are well above the average level. In fact, of all the Dice scores produced in the 10 experiments, only 20% have got ET Dice scores lower than the mean value, and in case of WT and TC Dice scores, it has been 26% and 29%, respectively. Thus, the mean scores represent more a small minority of the cases.

The consistency of the performance is also related to the reproducibility of the system in processing individual patient cases. For a given patient case, the system is expected to reproduce very similar results each time after the system is re-trained from the initial state. The results of 2 typical cases are presented in *Table 4.2*. The Dice scores obtained in the repeated training/testing processes are very consistent with standard deviations below 1% in most cases.

Table 4.1 Statistical Data of the Ten Experiments

| Exp. | Dice - Mean | | | Dice - Median | | | Dice - Mode | | |
|---|---|---|---|---|---|---|---|---|---|
| | ET | WT | TC | ET | WT | TC | ET | WT | TC |
| No. 1 | 0.769 | 0.892 | 0.759 | 0.858 | 0.919 | 0.874 | 0.89 | 0.93 | 0.95 |
| No. 2 | 0.758 | 0.888 | 0.754 | 0.855 | 0.915 | 0.863 | 0.89 | 0.93 | 0.93 |
| No. 3 | 0.782 | 0.893 | 0.768 | 0.847 | 0.919 | 0.861 | 0.85 | 0.93 | 0.93 |
| No. 4 | 0.764 | 0.889 | 0.767 | 0.853 | 0.918 | 0.865 | 0.89 | 0.93 | 0.95 |
| No. 5 | 0.790 | 0.894 | 0.769 | 0.863 | 0.918 | 0.858 | 0.89 | 0.95 | 0.95 |
| No. 6 | 0.768 | 0.893 | 0.771 | 0.857 | 0.920 | 0.847 | 0.89 | 0.95 | 0.95 |
| No. 7 | 0.763 | 0.888 | 0.755 | 0.850 | 0.922 | 0.857 | 0.89 | 0.93 | 0.93 |
| No. 8 | 0.779 | 0.897 | 0.765 | 0.851 | 0.917 | 0.852 | 0.85 | 0.93 | 0.93 |
| No. 9 | 0.783 | 0.894 | 0.764 | 0.854 | 0.916 | 0.858 | 0.89 | 0.95 | 0.95 |
| No. 10 | 0.766 | 0.893 | 0.761 | 0.859 | 0.922 | 0.848 | 0.91 | 0.93 | 0.93 |
| **Average** | **0.772** | **0.892** | **0.763** | **0.855** | **0.918** | **0.858** | **0.88** | **0.94** | **0.94** |
| **STDEV** | **0.010** | **0.003** | **0.006** | **0.005** | **0.002** | **0.008** | **0.02** | **0.01** | **0.01** |

Table 4.2 Dice Scores of 2 Patient Cases

| Exp. | Dice – Case 1 * | | | Dice – Case 2 ** | | |
|---|---|---|---|---|---|---|
| | ET | WT | TC | ET | WT | TC |
| No. 1 | 0.890 | 0.929 | 0.892 | 0.798 | 0.908 | 0.899 |
| No. 2 | 0.892 | 0.926 | 0.873 | 0.783 | 0.887 | 0.881 |
| No. 3 | 0.888 | 0.928 | 0.888 | 0.784 | 0.901 | 0.890 |
| No. 4 | 0.892 | 0.934 | 0.880 | 0.790 | 0.891 | 0.893 |
| No. 5 | 0.890 | 0.929 | 0.889 | 0.795 | 0.890 | 0.878 |
| No. 6 | 0.888 | 0.926 | 0.844 | 0.778 | 0.910 | 0.868 |
| No. 7 | 0.894 | 0.929 | 0.877 | 0.784 | 0.899 | 0.890 |
| No. 8 | 0.888 | 0.932 | 0.884 | 0.785 | 0.885 | 0.892 |
| No. 9 | 0.890 | 0.931 | 0.899 | 0.798 | 0.897 | 0.891 |
| No. 10 | 0.892 | 0.927 | 0.887 | 0.790 | 0.892 | 0.894 |
| **Average** | **0.890** | **0.929** | **0.881** | **0.789** | **0.896** | **0.888** |
| **STDEV** | **0.002** | **0.002** | **0.015** | **0.007** | **0.008** | **0.009** |

*Case 1: Brats18_MDA_922_1
**Case 2: Brats18_MDA_907_1

### 4.2.3 ABLATION STUDY

The proposed system is a custom-designed and involves an application-specific CNN for the brain tumor segmentation. All the blocks and the network elements/parameters are tailored to suit the input data to maximize the computation efficiency. A good number of trials, organized in 5 trial groups, have been conducted to assess their effects. Each Dice score presented in this subsection is the mean value resulted from 5~10 repeated training/testing processes.

The proposed CNN distinguishes itself by its ultra-simple structure, characterized by the 108 convolution kernels distributed in its 7 layers. In particular, the number of kernels per layer does not increase exponentially over the layers. The first trial group is to assess the computation efficiency of this structure. We created, from the proposed system, 3 variations by making the number of kernels in the convolution layers increase exponentially in the first half of the network, keeping the other parts unchanged. In Variation 1, the number of kernels grows from 16 to 64 over the first 3 layers and shrinks back to 16 in the last 3 layers. In case of the Variation 2, it is from 64 to 256. Variation 3 is created based on Variation 1, by replacing a single convolution by double ones. All of the 3 variations are U-net-like systems and Variation 3 bears more likeness to the U-net than the other 2.

The test results of the 4 systems, i.e., the proposed one and its 3 variations, are presented in *Table 4.3*. One can see that, in terms of processing quality, the results of the 4 systems are in the same level. However, to reach that level, the proposed system requires only a small fraction of the computation needed in these 3 U-net-like systems.

Increasing exponentially the kernel number, while filtering process goes "deeper", is conventional manner to cope with the complexity in feature data, but it is not necessarily effective in handling the feature information from brain images. The test results described above confirm that, in the proposed CNN, the number of kernels in each layer is appropriate for a correct transformation of the data originated from brain images. More kernels, requiring more computation in training and in testing, do not necessarily result in a significantly better processing quality, but putting unnecessary computation burden that may contribute adversely to the system in the aspects such as performance reliability, processing speed and system implementation.

Table 4.3 Results of Trial Group 1

| Systems | Dice | | | Number of trainable parameters | FLOPs per patient case |
| --- | --- | --- | --- | --- | --- |
| | ET | WT | TC | | |
| Variation 1 | 0.772 | 0.891 | 0.772 | 86.284 K | 53.51G |
| Variation 2 | 0.771 | 0.894 | 0.795 | 1363.492 K | 738.86G |
| Variation 3 | 0.769 | 0.892 | 0.789 | 177.172 K | 120.16G |
| Proposed system | 0.772 | 0.892 | 0.763 | 20.308 K | 29.07G |

The second trial group is to evaluate the effectiveness of the 2 convolution paths in the first 2 layers of the proposed CNN. For this purpose, a deviated version of the CNN is created by removing the depthwise convolution path while doubling the kernel number of the 2 standard convolutions. In other words, the deviated version doubles its capability to extract cross-modality features but there is no mono-modality feature extraction. The test results of this version are presented in *Table 4.4*, in comparison with those of the proposed CNN system. One can observe the following points.

- Even though the deviation, i.e., depthwise convolutions replaced by standard ones, changes only 50% of the first 2 layers, it results in a visible increase in the computation volume, measure by FLOPs and the number of parameters to be trained.
- Despite the increase in computation, the processing quality of the deviated system is not as good as the proposed one, which indicates that the mono-modality features, extracted by means of low-computation depthwise convolutions, is very useful in the process.

Table 4.4 Results of Trial Group 2

| Systems | Dice ET | Dice WT | TC | Number of trainable parameters | FLOPs per patient case |
|---|---|---|---|---|---|
| One standard-convolution pathway for the first two layers | 0.766 | 0.885 | 0.758 | 21.268K | 36.13G |
| Proposed system | 0.772 | 0.892 | 0.763 | 20.308K | 29.07G |

The third trial group is to measure the importance of the modality-wise normalization of the input data to the network. *Table 4.5* illustrates the test results obtained from the proposed system and its 2 modified versions. In the first modified version, the brain image slices are applied directly to the network without any normalization, and the results are presented in the first row. In the second modified version, the modality-wise normalization is replaced by a conventional batch normalization, and the results are presented in the second row. Compared them with the results of the proposed system, presented in the 3$^{rd}$ row, one can see that the proposed system with the modality-wise normalization applied to input data performs significantly better than the 2 other versions. It confirmed that it is critically important to normalize the input data before the first convolution and the modality-wise normalization is the most appropriate in case of 3D brain images.

Table 4.5 Results of Trial Group 3

| System with ... normalization before the first convolution | Dice ET | WT | TC |
|---|---|---|---|
| no normalization at all | 0.638 | 0.776 | 0.680 |
| batch normalization | 0.652 | 0.803 | 0.690 |
| modality-wise normalization | 0.772 | 0.892 | 0.763 |

The fourth trial group is to assess the efficiency of the proposed activation function Full-ReLU. To do so, we created another variation of the proposed system by putting a regular ReLU at the places of each Full-ReLU, meanwhile doubling the number of kernels in each the convolutions. The test results of the 2 systems are presented in *Table 4.6*.

Though the number of kernels in each of the convolution layers in the proposed system is only $\frac{1}{2}$ of that in the version without Full-ReLU, the processing quality of the 2 systems is at the same level whereas the difference in computation volume is very significant. They confirm that (i) Full-ReLU can effectively halve the number of kernels in a convolution layer of feature extraction, and (ii) doubling the kernels in the layers does not necessarily help to improve the processing quality.

Table 4.6 Results of Trial Group 4

| Systems | Dice ET | WT | TC | Number of trainable parameters | FLOPs per patient case |
|---|---|---|---|---|---|
| with regular ReLU, doubled number of kernels | 0.773 | 0.894 | 0.778 | 49.828K | 70.02G |
| Proposed system | 0.772 | 0.892 | 0.763 | 20.308K | 29.07G |

The proposed system consists of a pre-CNN block, CNN, and a post-CNN block. The fifth trial group is to assess the efficiency of the blocks and the results are found in *Table 4.7*. The Dice scores of the CNN stand-alone are presented in the first row, those of the pre-CNN combined with the CNN are found in the second row, and those in the third row are given by the complete system involving both pre- and post- CNN blocks. One can observe that the pre-CNN helps to reduce considerably the amount of computation, while helping to better detect enhancing tumors, whereas the post-CNN provides effectively a very noticeable improvement in the detection of enhancing tumors.

Table 4.7 Results of Trial Group 5

| Systems | Dice ET | Dice WT | Dice TC | Number of trainable parameters | FLOPs per patient case |
|---|---|---|---|---|---|
| Proposed CNN stand-alone | 0.722 | 0.891 | 0.763 | 20.308K | 35.20G* |
| Pre-CNN + proposed CNN | 0.731 | 0.891 | 0.763 | 20.308K | 29.07G |
| Complete proposed system | 0.772 | 0.892 | 0.763 | 20.308K | 29.07G |

* After removal of excessive margins in slices

In summary, the results of all the trials confirm a very high computation efficiency of the proposed system, i.e., having a high processing quality at a very low computation cost. In the setting of the proposed system, the number of kernels in each layer of the CNN is very small, but sufficient to process its input data, and there is no need to adopt conventional structure of exponentially growing layer width. This sufficiency is, however, in the context that all the network components are tailored to suit the input data and the task to secure the high performance. It has been proved that the use of Full-Relu helps to halve the number of kernels in layers of feature extraction, which contributes to the simplicity of the network. The trial results also confirm the importance of the data preparation by the modality-wise normalization to facility the processing in the convolution layers. The importance of mono-modality features, extracted by means of depthwise convolutions, has also been demonstrated. The contribution of the pre- and post- CNN to the improvement of the detection of enhancing tumors has also been confirmed. The trial results demonstrate that a CNN system for a specific application can be made very simple to achieve a high performance without need for a large amount of computation.

### 4.2.4 COMPARISON OF THE RESULTS

The test results of the proposed system are compared with those produced by seven other CNN systems having moderate network complexity and reported in recent years. The mean scores of Dice, Sensitivity, Specificity and Hausdorff95 of the proposed system, together with those of the seven systems, are presented in *Table 4.8*. The results of False Discovery Rate (*FDR*), False Negative Rate (*FNR*), and the measures of computation complexity, in terms of the number of trainable parameters and number of Flops per patient case, are found in *Table 4.9*. As training/testing samples play very important role in the performance evaluation, for each of the systems listed for comparison, the information about which datasets were used and whether the results were generated by CBICA Image Processing Portal is also found in the two tables.

It should be underlined that, the CNN block in the proposed system requires only 20308 parameters for its 7 convolution layers, as shown in *Table 4.9*. Its computation cost is significantly lower than those reported so far, and the system yields, nevertheless, a high processing quality. One can easily see that, compared with other brain tumor segmentation systems of modest computation, the proposed system has

- very good Dice scores,
- the lowest False Discovery Rates (*FDR*) in the detection of ET, WT and TC,
- False Negative Rates (*FNR*) or miss rates comparable to others, and
- the results among the best reported in the detection of ET voxels in the aspects of Dice and *FDR*. In particular, the *FDR* of ET is 13.8% lower than the second best found in the list.

The excellent processing quality is mainly owing to the specifically designed CNN for brain tumor segmentation. Though it has only 7 convolution layers, the operations in each layer are made to extract critical feature information, first for the localization of the tumor areas and then for the precise classification. Furthermore, the refinement block provides another improvement, after the CNN, in the ET detection: The average of Dice scores is increased from 73.1% to 77.2%, that of *FDR* reduced from 32.0% to 25.0% and Hausdorff95 reduced from 7.625 to 5.036.

Another important item in the performance metrics is the number of floating-point operations (FLOPs) required to complete the segmentation of each patient case. With a large amount of input data, i.e., 35.712M (240×240×155×4) voxels in 3D brain MRI images in each patient case, the proposed CNN requires only 29.07G FLOPs to complete the task. This extremely small number of FLOPs results mainly from the simplicity of the proposed CNN with 20308 parameters. Also, the pre-CNN block helps to reduce more than 50% of the input data volume applied to the CNN.

Table 4.8 Comparison of the Results – Dice, Sensitivity, Specificity and Hausdorff95

| Systems | Dataset | CBICA assessment | 2D/3D | Dice | | | Sensitivity | | | Specificity | | | Hausdorff95 | | |
|---|---|---|---|---|---|---|---|---|---|---|---|---|---|---|---|
| | | | | ET | WT | TC | ET | WT | TC | ET | WT | TC | ET | WT | TC |
| Chen et al. [13] | BRATS 2017 | No | 2D | 0.650 | 0.835 | 0.732 | 0.804 | 0.845 | 0.749 | 0.999 | 0.999 | 0.999 | 30.310 | 36.400 | 25.590 |
| Pereira et al. [14] | BRATS 2017 | **Yes** | 2D | 0.733 | 0.895 | 0.798 | N.A. | N.A. | N.A. | N.A. | N.A. | N.A. | 5.074 | 5.920 | 8.947 |
| Chen et al. [18] | BRATS 2017 | **Yes** | 3D | 0.703 | 0.891 | 0.782 | N.A. | N.A. | N.A. | N.A. | N.A. | N.A. | N.A. | N.A. | N.A. |
| Chen et al. [18] | BRATS 2017 | **Yes** | 3D | 0.735 | 0.893 | 0.739 | N.A. | N.A. | N.A. | N.A. | N.A. | N.A. | N.A. | N.A. | N.A. |
| Zhou et al. [19] | BRATS 2017 | No | 3D | 0.730 | 0.894 | 0.816 | N.A. | N.A. | N.A. | N.A. | N.A. | N.A. | 7.680 | 5.730 | 6.790 |
| Hu et al. [16] | **BRATS 2018** | **Yes** | 2D | 0.718 | 0.882 | 0.748 | 0.868 | 0.907 | 0.762 | 0.995 | 0.992 | 0.997 | 5.686 | 12.607 | 9.622 |
| Zhou et al. [17] | **BRATS 2018** | **Yes** | 3D | 0.753 | 0.864 | 0.774 | N.A. | N.A. | N.A. | N.A. | N.A. | N.A. | N.A. | N.A. | N.A. |
| **Proposed** | **BRATS 2018** | **Yes** | **2D** | **0.772** | **0.892** | **0.763** | **0.796** | **0.886** | **0.774** | **0.998** | **0.995** | **0.996** | **5.036** | **5.475** | **10.619** |

Table 4.9 Comparison of the Results – FDR, FNR, Computation Complexity/Volume

| Systems | Dataset | False Discovery Rate | | | False Negative Rate | | | 2D/3D | Max number of filters in a layer | Number of layers | Number of Parameters | Number of FLOPs per patient case |
| --- | --- | --- | --- | --- | --- | --- | --- | --- | --- | --- | --- | --- |
| | | ET | WT | TC | ET | WT | TC | | | | | |
| Chen et al. [13] | BRATS 2017 | 0.455 | 0.176 | 0.284 | 0.197 | 0.155 | 0.251 | 2D | N.A. | N.A. | 10.03M | N.A. |
| Pereira et al. [14] | BRATS 2017 | N.A. | N.A. | N.A. | N.A. | N.A. | N.A. | 2D | 160 | ≥ 12 | N.A. | N.A. |
| Chen et al. [18] | BRATS 2017 | N.A. | N.A. | N.A. | N.A. | N.A. | N.A. | 3D | 256 | 15 | 6.3M | N.A. |
| Chen et al. [18] | BRATS 2017 | N.A. | N.A. | N.A. | N.A. | N.A. | N.A. | 3D | 150 | 11 | 100K | N.A. |
| Zhou et al. [19] | BRATS 2017 | N.A. | N.A. | N.A. | N.A. | N.A. | N.A. | 3D | N.A. | ≥ 83 | Millions | N.A. |
| Hu et al. [16] | BRATS 2018 | 0.388 | 0.141 | 0.265 | 0.132 | 0.093 | 0.238 | 2D | 160 | ≥ 13 | N.A. | N.A. |
| Zhou et al. [17] | BRATS 2018 | N.A. | N.A. | N.A. | N.A. | N.A. | N.A. | 3D | 50 | ≥ 11 | Millions | N.A. |
| **Proposed** | **BRATS 2018** | **0.250** | **0.102** | **0.247** | **0.204** | **0.114** | **0.226** | **2D** | **24** | **7** | **20.308K** | **29.07G** |

## 5. CONCLUSION

In this paper, a computation-efficient CNN system has been proposed for brain tumor segmentation. It consists of three parts, a pre-CNN block to reduce the data volume, a unique CNN for the computation of segmentation, and a refinement block to detect false positive pixels. The CNN is custom-designed, following the new paradigm of ASCNN (application specific CNN) proposed in this paper. It has 7 convolution layers involving only 108 kernels, featuring (i) modality-wise normalization of the input data before the first convolution, (ii) new activation function Full-Relu halving the number of kernels in a convolution layer of high-pass filtering, (iii) 2-path convolution in the first 2 layers to extract mono-modality and cross-modality features, (iv) bilinear upsampling for dimension expansion without introducing randomness, and (v) weighted addition of the upsampled data and local feature data for signal modulation. In this specific design context, the number of kernels in each of the 7 layers is tailored to be just-sufficient according to the input of the layer and the task assigned to it, instead of exponentially growing over the layers, to increase information density and lower randomness in the processing.

The processing quality and the reproducibility of the proposed system has been evaluated with BRATS2018 dataset. The results demonstrate that the proposed system reproduces reliably almost the same output to the same input after retraining. The mean dice scores for enhancing tumor, whole tumor and tumor core are 77.2%, 89.2% and 76.3%, respectively. The system has 20308 trainable parameters and needs 29.07G Flops to test a case, i.e., a very small fraction of what is needed by the smallest CNN so far reported for the same segmentation.

It should be mentioned that the proposed system is not for general purpose. It has been designed to meet the specific needs to segment brain tumors or other kinds of tumors in medical images. In this way, the randomness and the redundancy in computation can be minimized, the information density in data flow increased, the dependency on training samples decreased, and the computation efficiency/quality improved.

This design demonstrates that a CNN system can be made to perform a high-quality processing at a very low computation cost for a specific application. Hence, application-specific CNN (ASCNN) is an effective approach to lowering the barrier of computation resource requirement of CNN systems to make them more implementable and applicable for general public.

## ACKNOWLEDGMENTS

This work was supported in part by Compute Canada and in part by the Natural Sciences and Engineering Research Council (NSERC) of Canada.